\documentclass[12pt, a4paper]{article}
\usepackage[T2A]{fontenc}
\usepackage[cp1251]{inputenc}
\usepackage{amsmath,amssymb}
\usepackage{amsfonts}
\usepackage{graphicx,color}
\usepackage{amsthm}

\graphicspath{{./Figs/}}
\DeclareGraphicsExtensions{.pdf,.jpg,.png,.eps,.gif}
\date{\today}
\newcommand{\EQ}{\begin{equation}}
\newcommand{\EN}{\end{equation}}

\newtheorem{theo}{Theorem}
\newtheorem{deff}{Definition}

\newtheorem{lem}{Lemma}

\newcommand{\pr}{\indent{\bf Proof. \ }}

\newcommand{\bx}{{\bf x}}

\newcommand{\F}{{\mathbb F}}
\newcommand{\PP}{{\mathbb P}}

\title{ On $q$-ary codes with two distances $d$ and $d+1$}

\author{P. Boyvalenkov$^*$, K. Delchev\footnote{Institute of Mathematics and Informatics,
Bulgarian Academy of Sciences, 8 G. Bonchev Str., 1113  Sofia, Bulgaria
(e-mail: peter@math.bas.bg, math\_k\_delchev@yahoo.com)},
D. V. Zinoviev$^\dagger$, V. A. Zinoviev\footnote{A.A. Kharkevich
Institute for Problems of Information Transmission, Russian
Academy of Sciences, Bol'shoi Karetnyi per. 19, GSP-4, Moscow,
101447, Russia (e-mail: \{dzinov,zinov\}@iitp.ru)}}

\date{}

\begin{document}
\maketitle

\begin{abstract}
The $q$-ary block codes with two distances $d$ and $d+1$ are
considered. Several constructions of such codes are given, as
in the linear case all codes can be obtained by a simple modification of
linear equidistant codes. Upper bounds for the maximum cardinality of
such codes is derived. Tables of lower and upper bounds for small
$q$ and $n$ are presented.
\end{abstract}

\section{Introduction}

Let $Q = \{0,1,...,q-1\}$. Any subset $C \subseteq Q^n$ is a code
denoted by $(n,N,d)_q$ of length $n$, cardinality $N = |C|$ and
the minimum (Hamming) distance $d$. For linear codes we use
notation $[n,k,d]_q$ (i.e., $N=q^k$). An $(n,N,d)_q$ code $C$ is
equidistant if for any two distinct codewords $x$ and $y$ we have
$d(x,y) = d$, where $d(x,y)$ is the (Hamming) distance between $x$
and $y$. A code $C$ is constant weight and denoted $(n,N,w,d)_q$
if every codeword is of weight $w$.

We consider codes with only two distances $d$ and $d+1$. As we will observe,
such codes are sometimes connected to equidistant codes. We are not aware, however,
of any investigations of codes with two consecutive distances.

Denote by $(n,N,\{d, d+1\})_q$ an $(n,N,d)_q$ code $C \subset Q^n$
with the following property: for any two distinct codewords $x$
and $y$ from $C$ we have $d(x,y) \in \{d, d+1\}$.  We are
interested in constructions, classification results and upper
bounds on the maximal possible size of $(n,N,\{d, d+1\})_q$ codes.
We show that the linear $q$-ary codes with two distances $d$ and $d+1$ are
completely known and can be obtained by simple modification of
linear equidistant codes [1,\,2]. The preliminary results of this
paper were announced partly in [1].

\section{Preliminary results}

We recall the following  classical Johnson bound  for the size $N_{q}(n, d, w)$
of a $q$-ary constant weight $(n,N,w,d)_q$-code [3]:
\begin{equation}
\label{eq:1.2}
N_{q}(n, d, w) \leq \frac {(q-1)dn}{qw^{2} -(q-1)(2w-d)n}
\end{equation}
if $qw^{2} > (q-1)(2w-d)n$.

\begin{deff}\label{deff:bib}
A balanced incomplete block (BIB) design $B(v,k,\lambda)$
is an incidence structure
$(X,B)$, where $X=\{x_1,\ldots,x_v\} $ is a set of elements
and $B$ is a collection of $k$-sets of elements (called {\em
blocks}) such that every two distinct elements of $X$ are contained in
exactly $\lambda \geq 0$ blocks of $B$ (here $1 \leq k \leq v-1$).
\end{deff}

Two other parameters of a $B(v,k,\lambda)$-design are $b = |B|$ (the number of blocks)
and $r$ (the number of blocks containing one fixed element):
\[
r = \lambda \,\frac{v-1}{k-1}\,,\;\;b = \lambda \,\frac{v(v-1)}{k(k-1)}\,,\;\;
\mbox{if}\;\;\lambda > 0;
\]
($\lambda = 0$ corresponds to the case $k=1$ and hence $b=rv$).

In terms of the binary incidence matrix a $B(v,k,\lambda)$-design is a binary
$(v \times b)$ matrix $A$ with columns of weight $k$ such that any
two distinct rows contain exactly $\lambda$ common nonzero positions.

We need the following result from [4, 5].
\begin{theo}\label{th:1}
Any $m$-nearly resolvable $NRB_m(v,k, \lambda)$-design induces a
$q$-ary equidistant constant weight $(n,N,w,d)_q$ code $C$ with the additional
property and with parameters $q=(v-m+k)/k$, $N=v$,
\[
n=\frac{\lambda v(v-1)}{(k-1)(v-m)},\;
w=\frac{\lambda(k-1)}{v-1},\;\;d=\frac{\lambda(v+m-k)}{k-1},
\]
meeting the Johnson bound (\ref{eq:1.2}).
\end{theo}

Recall the following wide class of $q$-ary equidistant codes constructed in [6].

\begin{theo}\label{th:2}
Let $p$ be a prime and let $s, \ell, h$ be any positive integers. Then there
exists an equidistant $(n,N,d)_q$ code with parameters
\[
q = p^{sh},\; n = \frac{p^{s(h+\ell)} - 1}{p^s-1},\;N = p^{s(h+\ell)}, \;
d = p^{s\ell} \cdot \frac{p^{sh}-1}{p^s-1}.
\]
\end{theo}

\begin{deff}\label{deff:dm}
Let $G$ be an abelian group of order $q$ written additively. A square
matrix $D$ of order $q\mu$ is called a difference matrix and denoted
$D(q, \mu)$, if the component-wise difference of any two different
rows of $D$ contains any element of $G$ exactly
$\mu$ times.
\end{deff}

Clearly a matrix $D(q, \mu)$ induces an equidistant
$(q\mu-1, q\mu, \mu(q-1))_q$ code [6].

\section{Constructions}

\subsection{Combinatorial constructions}

Denote by $W_q(n)$ a ball of radius $1$ with center at the zero
vector, i.e. $ W_q(n) = \{x \in Q^n: \,\mbox{wt}(x) \leq 1\}$.

{\em Construction 1a.} The ball $W_q(n)$ is an
$(n,(q-1)n+1,\{1,2\})_q$ code.

{\em Construction 1b.} Parity checking (modulo 2) of
Construction 1a implies an $(n+1,(q-1)n+1,\{2,3\})_q$
code, which we denote by $W^*_q(n+1)$. For any codeword $(0 \cdots
0~a~0 \cdots 0)$ from $W_q(n)$ we form the codeword $(0 \cdots
0~a~0 \cdots 0~|~a)$ from $W^*_q(n+1)$.

{\em Construction 2.} An equidistant $(n,N,d)_q$ code $C$ produces
two $(n',N,\{d', d'+1\})_q$ codes, namely, $(n-1,N,\{d-1, d\})_q$
code $C_1$ obtained by deleting (any) position from $C$, and
$(n+1,N,\{d, d+1\})_q$ code $C_2$ obtained by adding one position
to $C$.

Combining Constructions 1a and 1b with Construction 2 we obtain the following
two constructions.

{\em Construction 3a.} An equidistant
$(n_1,N_1,d)_{q_1}$ code and $W_{q_2}(n_2)=(n_2,N_2,\{1,2\})$ give
an $(n,N,\{d+1,d+2\})_q$ code with parameters
\[
q = \max\{q_1,q_2\},\;\;n = n_1 + n_2,\;\;N = \min\{N_1, N_2\}.
\]

{\em Construction 3b.} An equidistant
$(n_1,N_1,d)_{q_1}$ code and $W_{q_2}^*(n_2)=(n_2,N_2,\{2,3\})$
give an $(n,N,\{d+2,
d+3\})_q$ code with parameters
\[
q = \max\{q_1,q_2\},\;\;n = n_1 + n_2,\;\;N = \min\{N_1, N_2\}.
\]

{\em Construction 4.} If there exist $r$ mutually orthogonal Latin
squares of order $q$, then there exists a family
of $(s+2,q^2,\{s+1, s+2\})_q$ codes $C_s$, where $s = 1, \ldots,
r$.

Combining Constructions 2 and 4, we obtain:

{\em Construction 5.} For any prime power $q$ there exists a
family of $(n,q^2,\{d, d+1\})_q$ codes with parameters
\[
n = s(q+1) + r,\;\;d = sq + r-1,\;\;s \geq 1,\;\;r = 1, \ldots, q+1.
\]

{\em Construction 6.} If there exists a difference matrix $D(q,\mu)$,
then there exist $(n,N,\{d,d+1\})_q$ codes with parameters:
\[
\begin{array}{ccl}
n = q\mu-2,\;&\;N = q\mu,\;&\;d = (q-1)\mu-1,\\
n = q\mu,\;&\;N = q\mu,\;&\;d = (q-1)\mu.\\
\end{array}
\]

The well known equidistant $(4,9,3)_3$ code $C_1$ and a
$(4,9,\{1,2\})_3$ code $C_2$ (Construction 1a) give by
Construction 3 an $(8,9,\{4,5\})$ code $C$, which is not good.
Using the $(5,9,\{2,3\})_3$ code $C_3$ (Construction 1b) gives by
Construction 3 a $(9,9,\{5,6\})$ code.

The equidistant $(13,27,9)_3$ (Theorem \ref{th:2}) implies by
Construction 2 a $(14,27,\{9,10\})_3$ code, which is better than
the random $(14,18,\{9,10\})_3$ code and also a
$(12,27,\{8,9\})_3$ code which meets the upper bound (the best found random
code has cardinality $18$).

The difference matrix $D(4,3)$ (see [7]) without the
trivial column is an optimal equidistant $(11,12,8)_3$ code.
The difference matrix $D(3,4)$ (see [7]) without the
trivial column is an equidistant $(11,12,9)_4$ code.

The well known equidistant $(5,16,4)_4$ code $C_1$ and a
$(5,16,\{1,2\})_4$ code $C_2$ (Construction 1) give by
Construction 3 a  $(10,16,\{5,6\})_4$ code (not good --
there is a random $(10,20,\{5,6\})_4$ code). Twofold
repetition of $(5,16,4)_4$ code $C_1$ gives an optimal
$(10,16,8)_4$ code.

The equidistant $(6,9,5)_4$ code [5] implies by twofold
repetition a $(12,9,\{10,11\})_4$ code (better than the random
code). The equidistant $(21,64,16)_4$ code [6] implies
$(22,64,\{16,17\})_4$ and $(20,64,\{15,16\})_4$ codes by
Construction 2.
The equidistant $(9,10,8)_5$ code [4] implies
$(8,10,\{7,8\})_5$ and $(10,10,\{8,9\})_5$ codes by Construction
2. By Construction 5 we obtain the following family of
$(n,N,\{d,d+1\})_5$ codes:
\[ n = 9 + s,\;\; N = 10,\;\; d = 8 + s-1,\;\; s = 0, 1, \ldots, 6. \]
In particular, for $s=0$ we obtain an optimal $(9,10,8)_5$ code
and for $s \geq 2$ all resulting codes are new. By Construction 1
this equidistant $(9,10,8)_5$ code implies the $(11,9,\{9,10\})_5$
code.

The equidistant $(6,25,5)_5$ code implies the family of
$(6+s,25,\{5+s-1,5+s\})_5$ codes where $s = 0,1, \ldots, 6$, which
give better (or new codes) for $s \geq 1$.

The well known resolvable design $(15,35,7,3,1)$ is equivalent to
the optimal equidistant $(7,15,6)_5$ code. Now using Construction
5 we obtain from this code the following codes: $
n = 7 + s,\;\; N = 15,\;\; d = 6 + s - 1,\;\;s = 1, \ldots, 6. $

The affine design $(16,20,5,4,1)$ implies [4] the
equidistant constant weight $(16,16,15,14)_6$ code which implies
in turn the $(16,17,\{14, 15\})_6$ code (by adding the zero
codeword).

\subsection{Random codes}

We use a computer program for generation of random codes by a simple heuristic
algorithm. We start with a seed (at least the zero vector), then
generate the search space and choose consecutively random vectors
until the resulting code is good (i.e. has only distances $d$ and
$d+1$). It is possible to take for a seed the best code constructed
earlier. Many iterations can be implemented but usually the best codes (found this way) are obtained quickly.
The cardinalities of such random codes are shown in Section \ref{tables} together
with those of the codes obtained from constructions from this section.

\section{Linear $(n,N,\{d, d+1\})_q$ codes}

In this section we obtain the classification results in the case of linear
codes with distances $d$ and $d+1$. As we already mentioned the linear
codes with two distances are completely known. The next theorem was
proved for the binary case in [1]. The $q$-ary case also was conjectured
in [1]. Here we give a simple proof of our conjecture for the case
$q \geq 2$ and $k \geq 2$ based on purely
coding theoretic arguments.  Simultaneously
the corresponding result for $k \geq 3$ was proved in [2],
based on geometrical arguments.

Let $C$ be a $q$-ary (linear) $[n,3,q^2]_q$ equidistant code of
length $n = q^2+q+1$, the  distance $q^2$ and cardinality $q^3$.

\begin{lem}\label{lem:1}
Suppose that $C$ is the code above presented as a $(n \times q^3)$-matrix
over $\F_q$ (which we denote by $[C]$). Then, $[C]$ cannot be written as a
concatenation of two matrices, i.e. $[C] = [C_1 | C_2]$, where $[C_1]$ is
a $(x \times q^3)$-matrix (where $x < (n-1)/2$) which represents a linear
$(x,q^3,\{d,d+1\})_q$ code $C_1$.
\end{lem}

\pr
If $\bx\in C$ then clearly $\alpha\bx\in C$ for all $\alpha\in\F_q^*$.
Thus, all non zero codewords of $C$ can be split into $q^2+q+1$
classes. So, by $\PP C$ we denote the code of classes of such elements.
It is given by a matrix $P$ of $n$ by $n$, where $n=q^2+q+1$.

Suppose that the matrix $P$ is a concatenation of two matrices $P_1$ and $P_2$, i.e.
$P = [P_1|P_2]$, where $P_1$ is a $x$ by $q^2+q+1$ matrix that corresponds
to the equivalence classes of a $[x,q^3,\{d,d+1\}]$ linear code $C_1$.
Consequently every word of $C_1$ has $x-d$ or $x-d-1$ positions with zero
entries. To simplify further computations, let $\ell = x-(d+1)$ (since we
will consider the number of zero entries of any word instead of its weight).

Since $P$ corresponds to an equidistant code with code distance $q^2$,
clearly $P_2$ corresponds to a (linear) $[n-x,3,q^2-d-1]_q$ code $C_2$
with two distances $q^2-d-1$ and $q^2-d$. Therefore,
without loss of generality we can assume that $x \leq n/2$.
Since $n = q^2 + q + 1$, we assume that $x \leq q(q+1)/2$ and $\ell \leq (q+1)/2$
(indeed, every column of $P$ contains $q+1$ zeroes).

Since any column in $P$ has $q+1$ zeroes, the matrix $P$ contains
$n(q+1)$ zero entries. Suppose that the matrix $P_1$ has exactly $\eta$ words of weight
$d+1$ (i.e. $\ell$ zeroes) and the remaining $n-\eta$ words of weight $d$ (i.e. $\ell+1$
zeroes). Thus, we can write
$$
\ell \eta + (\ell+1)(n-\eta) = (q+1)x.
$$
Solving for $\eta$, we obtain
\EQ\label{eq:equation1}
\eta = (l+1)n - (q+1)x.
\EN
Since $C_1$ is a linear of dimension $3$, for any pair of coordinate positions
there exists exactly one row in $P_1$ with zeroes at these positions.
There are $x$ coordinate positions and $x(x-1)$ pairs of positions.
On the other hand, there are $\eta$ rows with $\ell$ zeroes
(every row provides $\ell(\ell-1)/2$ pairs of coordinates)
and $n-\eta$ rows with $\ell+1$ (every row provides $\ell(\ell+1)/2$ pairs of
coordinates). Thus, we obtain the following equality:
\EQ\label{eq:equation2}
\frac{x(x-1)}{2} = \frac{\ell(\ell-1)}{2} \eta + \frac{\ell(\ell+1)}{2}(n-\eta).
\EN
Our goal is to show that the equality (\ref{eq:equation2}) can not be valid for
any $x$ in the interval $[2,q(q+1)/2]$.
Using (\ref{eq:equation1}), the expression (\ref{eq:equation2}) becomes
\begin{eqnarray*}
x(x-1)
& = & \ell(\ell-1) \eta + \ell(\ell+1)n - \ell(\ell+1) \eta\\
& = & \ell(\ell+1)n - 2\ell \eta\\
& = & \ell(\ell+1)n - 2\ell[(\ell+1)n - (q+1)x]\\
& = & 2\ell(q+1)x - \ell(\ell+1)n.
\end{eqnarray*}
Thus, we arrive at the following quadratic equation for $x$:
\EQ\label{eq:equation3}
x^2 - (2\ell(q+1)+1)x + \ell(\ell+1)n = 0.
\EN
We will show that the discriminant of this equation is negative.
Thus, we have to verify that
$$
(2\ell(q+1)+1)^2 < 4\ell(\ell+1)n.
$$
Recalling that $n=q^2+q+1$, this is equivalent to
\begin{eqnarray*}
4\ell^2(q^2+2q+1) + 4\ell(q+1) + 1
& < & 4\ell(\ell+1)(q^2 + q + 1)\\
& = & 4\ell^2(q^2 + q + 1) + 4\ell(q^2+q+1).
\end{eqnarray*}
Once simplified, it becomes
$$
4\ell^2 q + 1 < 4\ell q^2.
$$
Since $\ell \leq (q+1)/2$, the last inequality is obviously true for $\ell\geq 1$.
Thuw, we obtained that there is no submatrix $P_1$, and consequently,
the linear $[q^2+q+1,3,q^2]_q$ code $C$ cannot be presented as a concatenation
of two linear codes $C_1$ and $C_2$ of type $(n,N,\{d,d+1\})_q$.
\qed

\begin{theo}\label{th:3}
Let $C$ be a $q$-ary linear $[n,k,d]_q$ code with two distances
$d$ and $d+1$ and $k \geq 2$. Then $C$ is obtained by Construction 2
from the previous section, i.e. by deleting or adding an arbitrary vector
column in the parity check matrix of a linear $q$-ary equidistant code with the following exception
for the case $k=2$ and $q \geq 3$, when $C$ can be obtained by
Construction 2 or by Construction 5.
\end{theo}

\pr
First consider the case $k=2$. For this case we can have a $[n,2,d]_q$ code
$C$ with two distances $d$ and $d+1$ obtained also by Construction 5.
Let $C_1$ be an equidistant $[n_1,2,d_1]_q$ code with parameters $n_1=s(q+1)$, \,
$d_1=s\,q$ and $C_2$ be a $[n_2,2,d_2]_q$ code with parameters $n_2=r$, \,
$d_2=r-1$. The generator matrix $G$ of $C$ is of the form $G = [G_1\,|\,G_2]$,
where $G_1$ and $G_2$ are the generator matrices of the codes $C_1$ and $C_2$,
which (up to equivalence) look as follows: the matrix $G_1 = [G_0| \cdots |G_0]$
is the $s$-time repetition of $G_0$,
\[
G_0 = \left[
\begin{array}{cccc}
a_0\;a_1\;a_1\;a_1&\;\cdots\,&a_1\,&\\
a_1\;a_0\;a_1\;a_2&\;\cdots\,&a_{q-1}&\\
\end{array}
\right]
\]
where we denote $\F_q = \{a_0=0,a_1=1, a_2, \ldots , a_{q-1}\}$
and the matrix $G_2$ is of the form
\[
G_2 = \left[
\begin{array}{cccc}
a_0\;a_1\;a_1\;a_1&\;\cdots\,&a_1\,&\\
a_1\;a_0\;a_1\;a_2&\;\cdots\,&a_{r-2}&\\
\end{array}
\right].
\]
All these facts are commonly known and do not need any proofs. The only thing
we have to say is that the all elements of the second row of $G_2$ starting
from the second position should be different and this condition is
necessary and sufficient in order for $G_2$ to be a generator matrix of the code $C_2$.

Now we claim that any $[n,2,d]_q$ code with two distances should be of
the form described above. It is clear for the case $n \leq q$. For larger
$n$ assume that the code $C_1$ of length $q+1$ is not equdistant $[q+1,2,q]_q$
code, i.e. it has minimal distance $d=q-1$. Since its averige distance is known
(and it equals $q$), we conclude that this code has three distances,
namely,  $q-1$, $q$ and $q+1$. Denoting by $\alpha_w$ the number of codewords
of weight $w$, and taking into account that $\alpha_{q-1} = \alpha_{q+1}$,
we obtain that
\EQ\label{eq:3.1}
\alpha_{q-1} = \alpha_{q+1} = q-1,\;\;\alpha_q = (q-1)^2.
\EN
As we know the $[r,2,r-1]_q$ code $C_2$ has weights $r-1$ and $r$. Denoting
$\beta_w$ the number of codewords of weight $w$, we deduce that
\EQ\label{eq:3.2}
\beta_{r-1} = (q-1)r,\;\;\beta_r = (q-1)(q+1-r).
\EN
So, the concatenation of these two codes $C_1$ and $C_2$ would be a code
$C$ with at least three distances $d, d+1$ and $d+2$ where $d \leq q+r-1$,
i.e. we obtain a contradiction. So, $C_1$ of length $(q+1)s$ should be an
equidistant code.
Therefore, any $[n,2,d]_q$ code $C$ with two distances $d, d+1$ is
obtained by one of two constructions, namely, Constructions 2 or 5.

Now to finish the proof we have only to show that any $[n,3,d]_q$-code
with two distances $d$ and $d+1$ can be obtained only by Construction 2.
In contrary, assume that $C_1$ is a $[n_1,3,d_1]_q$-code
with two distances $d_1$ and $d_1+1$ of length $n_1$ in the interval
$2 \leq n_1 \leq q^2+q-1$. It means that there is a $[n_2,3,d_2]_q$-code
$C_2$ (complementary to $C_1$) with two distances $d_2$ and $d_2+1$
of length $n_2 = q^2 + q + 1 - n_1$.
Hence, there exists a $q$-ary $[n,3,q^2]_q$ equidistant code $C$ of
length $n = q^2+q+1$, which can be written as a
concatenation of two codes $C_1$ and $C_2$.
But by Lemma \ref{lem:1} it is impossible, that finishes the proof.
\qed

\section{Upper bounds}

We are interested in upper bounds for the quantity
\[ A_q(n;\{d,d+1\}) = \max \{ |C| : C \mbox{ is an $(n,|C|, \{d,d+1\})$ code}\}, \]
the maximal possible cardinality of a code in $Q^n$ with two distances $d$ and $d+1$.

\subsection{Linear programming bounds}

For fixed $n$ and $q$, the (normalized) Krawtchouk polynomials are
    defined by
    \[
    Q_i^{(n,q)}(t) :=\frac{1}{r_i} K_i^{(n,q)}(d), \ d=\frac{n(1-t)}{2},  \ r_i=(q-1)^i {n \choose i},
    \]
    where
\[ K_i^{(n,q)}(d)=\sum_{j=0}^{i}   (-1)^j(q-1)^{i-j} {d \choose j} {n-d \choose i-j} \]
    are the (usual) Krawtchouk polynomials. If $f(t) \in \mathbb{R}[t]$ is of degree $m \geq 0$, then it can
    be uniquely expanded as
\[ f(t) =    \sum_{i=0}^m f_i Q_i^{(n,q)}(t). \]
The next theorem is adapted for estimation of $A_q(n;\{d,d+1\})$ from the general Delsarte linear programming bound .
Proofs of such bounds are usually considered folklore.

    \begin{theo}
        \label{thm lp}
        Let $n \geq q \geq 2$ and $f(t)$ be a real
        polynomial of degree $m \leq n$ such that:

        {\rm (A1)} $f(t) \leq 0$ for $t \in  \{1-2d/n,1-2(d+1)/n\}$;

        {\rm (A2)} the coefficients in the Krawtchouk expansion $f(t) =
        \sum_{i=0}^{m} f_i Q_i^{(n,q)}(t)$ satisfy $f_i \geq 0$ for every $i$.

        Then $A_q(n;\{d,d+1\}) \leq f(1)/f_0$.
    \end{theo}

Most of the upper bounds in the table below are obtained by Theorem \ref{thm lp} with the simplex method.
We describe now other cases where analytic forms of good bounds are possible.

The first degree polynomial $f(t)=t-1+2d/n$ gives the Plotkin bound which is attained for many large $d$.
Optimization over the second degree polynomials gives the following result.

\begin{theo}
        \label{thm lp1}
If $d \geq (n-1)(q-1)/q$, then
\begin{eqnarray}
\label{d2-bound}
A_q(n;\{d,d+1\}) \leq  \frac{q^2d(d+1)}{n^2(q-1)^2-n(q-1)(2dq+q-1)+dq^2(d+1)}.
\end{eqnarray}

    \end{theo}

\pr Consider the second degree polynomial
\[  f(t) = \left(t-1+\frac{2d}{n}\right)\left(t-1+\frac{2d+2}{n}\right) = f_0 +f_1 Q_1^{(n,q)}(t)+f_2 Q_2^{(n,q)}(t), \]
where $f_0 = \frac{4(n^2(q-1)^2-n(q-1)(2dq+q-1)+dq^2(d+1))}{n^2q^2}$, $f_1 = \frac{8(q-1)(dq-(q-1)(n-1))}{nq^2}$, and $f_2 = \frac{4(q-1)^2(n-1)}{nq^2}$. The condition (A1) is obviously satisfied.

The condition $f_0>0$ is equivalent to a quadratic inequality with respect to $dq$, giving that $n \geq q$ implies it.
The condition $f_1 \geq 0$ is equivalent to $dq \geq (n-1)(q-1)$ and $f_2>0$ is obvious. Thus $f(t)$ satisfies (A1) and (A2) provided
$d \geq (n-1)(q-1)/q$.

Now the calculation of $f(1)/f_0$ gives the bound \eqref{d2-bound}.
\qed

The bound \eqref{d2-bound} is attained in some cases. It gives
$A_q(n;\{d,d+1\}) \leq q^2$ for $d=n-1$ which is attained for $(q,n)=(3,3), (3,4), (4,5)$, and $(5,6)$.
Further, we have $A_2(7;\{4,5\})=A_2(7;\{3,4\})=8$, $A_2(10;\{5,6\})=12$, $A_3(12,\{8,9\})=A_3(13,\{9,10\})=27$
by \eqref{d2-bound}.
The cases of attaining \eqref{d2-bound} are marked by $d2$ in the tables below.

Furthermore, the bound \eqref{d2-bound} is attained by some code $C$ and $d>(n-1)(q-1)/q$ (i.e. $f_1>0$), then $C$ is
an orthogonal array of strength 2. In particular, the cardinality of $C$ is divisible by $q^2$. This argument implies improvements
of \eqref{d2-bound} by one giving the exact values $A_2(12,\{5,6\})=A_2(12,\{6,7\})=A_2(13,\{6,7\})=13$ and
the bounds $13 \leq A_3(6,\{4,5\}) \leq 14$. These
cases are marked by $n$ in the tables. One more interesting case is $A_3(7,\{4,5\})=15$ where \eqref{d2-bound}
is attaned in the case $d=(n-1)(q-1)/q$.

Further bounds can be obtained by some ad-hoc polynomials. For example, the polynomial
\[ f(t)=1+(q-1)nQ_{(n(q-1)+1)/q}^{(n,q)}(t) \]
gives $A_q(n,\{1,2\})=(q-1)n+1$ (see Construction 1a) whenever $q$ divides $(q-1)n+1$.
Similarly, the polynomial
\[ f(t)=1+\frac{n+2}{2}Q_{n/2}^{(n,2)}(t)+\frac{n}{2}Q_{1+n/2}^{(n,2)}(t), \]
where $n$ is even, gives $A_2(n,\{1,2\}) \leq f(1)/f_0=n+2$. This bound cannot be attained since it implies
impossible distance distributions for the corresponding codes. Both polynomials prove that
$A_2(n,\{1,2\})=n+1$ (attained by Construction 1a). Such cases are marked with $a$ in the tables.

Further careful examination of the conditions for attaining the linear programming bounds could probably lead to
other improvements in the tables.

\subsection{Bounds via spherical codes}

Codes from $Q^n$ are naturally mapped to the sphere $\mathbb{S}^{(q-1)n-1}$.
We first map bijectively the alphabet symbols $0,1,\ldots,q-1$ to the vertices of the regular simplex in $q-1$ dimensions
and then map the codewords of a $q$-ary code $C \subset Q^n$ coordinate-wise to $\mathbb{R}^{(q-1)n}$. It is easy to see that all vectors have the same norm
and after a normalization we obtain a spherical code on $\mathbb{S}^{(q-1)n-1}$. This spherical code has
cardinality $|C|$ and maximal inner product $1-2dq/(q-1)n$ (equivalently, squared minimum distance $2dq/(q-1)n$).
Clearly, $q$-ary codes with distances $d$ and $d+1$ are mapped to spherical 2-distance codes with
squared distances $2dq/(q-1)n$ and $2(d+1)q/(q-1)n$. This relation implies the following upper bound for
$A_q(n,\{d,d+1\})$.

\begin{theo}
        \label{thm sc}
If $d>(\sqrt{2(q-1)n}-1)/2$, then
\[ A_q(n,\{d,d+1\}) \leq 2(q-1)n+1. \]
    \end{theo}

\pr Larman, Rogers, and Seidel [11] proved that if the cardinality of a two-distance set in $\mathbb{R}^n$
with distances $a$ and $b$, $a <b$, is greater than $2n + 3$, then the ratio $a^2/b^2$ equals $(k-1)/k$,
where $k$ is
a positive integer satisfying $2 \leq k \leq (\sqrt{2n}+1)/2$. The restriction $2n+3$ was moved to $2n+1$ in
[12].

In our situation $a^2/b^2=d/(d+1)=(k-1)/k$ holds, whence we conclude that $d=k-1$ has to belong to the
interval $[1,(\sqrt{2(q-1)n}-1)/2]$. In other words, there exist no $q$-ary codes with distances $d$ and $d+1$
and cardinality greater than $2(q-1)n+1$, whenever $d>(\sqrt{2(q-1)n}-1)/2$; i.e., we have $A_q(n,\{d,d+1\}) \leq 2(q-1)n+1$.
\qed

The bound from Theorem \ref{thm sc} is usually better than the simplex method for large enough $n$ and middle range $d$.
The first time where this happens is $(n,d)=(13,4)$ for $q=2$, $(9,3$ for $q=3$, $(8,3)$ for $q=4$, and $(7,4)$ for $q=5$.

\section{Tables}
\label{tables}

The tables below are for $q=2,3,4,5$. Horizontally we give $d$, vertically $n$. The lower bounds show the better of the computer generated random
codes and the constructions from Section 3. All our random codes are available upon request.

The upper bounds are taken from the best of the linear programming bound obtained by the simplex method (unmarked),
ad-hoc approaches as in Section 5.1 (marked with $d2$, $n$ and $a$, respectively), the corresponding best known
upper bound on $A_q(n,d)$ [8] (marked with $*$), and the bound from Theorem \ref{thm sc} (marked with $t6$).
\begin{center}
{\scriptsize
\noindent
\begin{tabular}{|c|c|c|c|c|c|c|c|c|c|c|c|c|c|c|c|c|c|c|}
\hline
\multicolumn{18}{|c|}{$q=2$} \\
\hline
 & 1 & 2 & 3 & 4 & 5 & 6 & 7 & 8 & 9 & 10 & 11 & 12 & 13 & 14 & 15 & 16 & 17 \\
\hline
7 & 8 & 7-10 & $8^{d2}$ & $8^{d2}$ & 2* & 2* & & & & & & & & & & & \\ \hline
8 & 9$^a$ & 8-12 & 8-10 & 8-10 & 4* & 2* & 2* & & & & & & & & & & \\
\hline
9 & 10 & 9-14 & 8-16 & 8-10 & 6* & 4* & 2* & 2* & & & & & & & & & \\
\hline
10 & 11$^a$ & 10-16 & 8-16 & 10-16 & $12^{d2}$ & $6^{d2}$ & 2-3 & 2* & 2* & & & & & & & & \\
\hline
11 & 12 & 11-18 & 8-19 & 10-20 & $12^{d2}$ & $12^{d2}$ & 4* & 2* & 2* & 2* & & & & & & & \\
\hline
12 & 13$^a$ & 12-20 & 8-25 & 10-21 & $13^{n}$ & $13^{n}$ & 4* & 4* & 2* & 2* & 2* & & & & & & \\
\hline
13 & 14 & 13-22 & 8-26 & 10-27 & 13-19 & $13^{n}$ & $8^{d2}$ & 4* & 2* & 2* & 2* & 2* & & & & & \\
\hline
14 & 15$^a$ & 14-24 & 8-29$^{t6}$ & 10-29$^{t6}$ & 14-27 & 14-19 & $16^{d2}$ & 8* & 4* & 2-3 & 2* & 2* & 2* & & & & \\
\hline
15 & 16 & 15-26 & 8-31$^{t6}$  & 11-31$^{t6}$  & 14-29 & 14-30 & 16 & 16* & 4* & 4* & 2* & 2* & 2* & 2* & & & \\
\hline
16 & 17$^a$ & 16-28 & 8-33$^{t6}$  & 11-33$^{t6}$  & 14-33$^{t6}$  & 15-33$^{t6}$  & 16-18 & 16-18 & 6* & 4* & 2* & 2* & 2* & 2* & 2* & & \\
\hline
17 & 18 & 17-30 & 9-35$^{t6}$  & 12-35$^{t6}$  & 14-35$^{t6}$  & 15-35$^{t6}$  & 17-22 & 16-18 & 10* & 6* & 4* & 2* & 2* & 2* & 2* & 2* & \\
\hline
18 & 19$^a$ & 18-32 & 9-37$^{t6}$  & 12-37$^{t6}$  & 14-37$^{t6}$ & 15-37$^{t6}$  & 17-35 & 18-22 & 20* & 10 & 4* & 2-4 & 2* & 2* & 2* & 2* & 2* \\
\hline
\end{tabular}
}
\end{center}

\begin{center}
{\scriptsize
\noindent
\begin{tabular}{|c|c|c|c|c|c|c|c|c|c|c|c|c|c|}
\hline
\multicolumn{14}{|c|}{$q=3$} \\ \hline
 & 1 & 2 & 3 & 4 & 5 & 6 & 7 & 8 & 9 & 10 & 11 & 12 & 13 \\
\hline
3 & 9 & 9 & & & & & & & & & & & \\
\hline
4 & 9$^a$ & 9 & 9 & & & & & & & & & & \\
\hline
5 & 11-13 & 9-17 & 11-13 & 6 & & & & & & & & & \\
\hline
6 & 13-15 & 11-18 & 11-16 & 13-$14^n$ & 4 & & & & & & & & \\
\hline
7 & 15$^a$ & 13-27 & 11-27 & $15^{d2}$ & 10 & 3 & & & & & & & \\
\hline
8 & 17-19 & 15-31 & 11-30 & 15-31 & 18-19 & 9 & 3 &  & & & & & \\
\hline
9 & 19-21 & 17-33 & 11-37$^{t6}$ & 15-36 & 18-25 & 18-21 & 6 & 3 &  & & & & \\
\hline
10 & 21$^a$ & 19-45 & 11-41$^{t6}$ & 15-41$^{t6}$ & 18-41$^{t6}$ & 18-21 & 13-14 & 3 & & & & & \\
\hline
11 & 23-25 & 21-45 & 11-45$^{t6}$ & 15-45$^{t6}$ & 18-45 & 18-45 & 18-25 & $12^{*}$ & $4^{*}$ & $3^{*}$ & & & \\
\hline
12 & 25-27 & 23-51 & 12-49$^{t6}$ & 15-49$^{t6}$ & 18-49$^{t6}$ & 18-49$^{t6}$ & 18-30 & $27^{d2}$ & $9^{*}$ & $4^{*}$ & $3^{*}$ &  & \\
\hline
13 & 27$^a$ & 25-63 & 13-53$^{t6}$ & 15-53$^{t6}$ & 18-53$^{t6}$ & 18-53$^{t6}$ & 18-53$^{t6}$ & 18-27 & $27^{d2}$ & $6^{*}$ & $3^{*}$ & $3^{*}$ & \\
\hline
14 & 29-31 & 27-63 & 14-57$^{t6}$ & 15-57$^{t6}$ & 18-57$^{t6}$ & 18-57$^{t6}$ & 18-57$^{t6}$ & 18-45 & 27-31 & 12-13 & $6^{*}$ & $3^{*}$ & $3^{*}$ \\
\hline
\end{tabular}
}
\end{center}

\begin{center}
{\scriptsize
\noindent
\begin{tabular}{|c|c|c|c|c|c|c|c|c|c|c|c|}
\hline
\multicolumn{12}{|c|}{$q=4$} \\
\hline
 & 1 & 2 & 3 & 4 & 5 & 6 & 7 & 8 & 9 & 10 & 11 \\
\hline
5 & 16$^a$ & 16-25 & 16 & 16$^{*}$ & & & & & & & \\
\hline
6 & 19-22 & 16-37 & 16-37 & 18-22 & 9$^{*}$ & & & & & &\\
\hline
7 & 22-26 & 19-41 & 16-43 & 18-41 & 21-26 & 8$^{*}$ & & & & &\\
\hline
8 & 25-28 & 22-50 & 16-49$^{t6}$ & 18-49$^{t6}$ & 21-32 & 19-28 & 5$^{*}$ & & & & \\
\hline
9 & 28$^a$ & 25-67 & 16-86 & 18-55$^{t6}$ & 21-55$^{t6}$ & 19-28 & 15-20$^{*}$ & 5$^{*}$ & & &\\
\hline
10 & 31-34 & 28-72 & 16-90 & 18-61$^{t6}$ & 20-61$^{t6}$ & 19-61$^{t6}$ & 21-34 & 16$^{*}$ & 5$^{*}$ & &\\
\hline
11 & 34-38 & 31-78 & 16-134 & 18-67$^{t6}$ & 21-67$^{t6}$ & 19-67$^{t6}$ & 20-56 & 22-38 & 12$^{*}$ & 4$^{*}$ & \\
\hline
12 & 37-40 & 34-97 & 18-152 & 18-73$^{t6}$ & 21-73$^{t6}$ & 19-73$^{t6}$ & 20-73$^{t6}$ & 22-43 & 21-40 & 9$^{*}$ & 4$^{*}$  \\
\hline
\end{tabular}
}
\end{center}

\begin{center}
{\scriptsize
\noindent
\begin{tabular}{|c|c|c|c|c|c|c|c|c|c|}
\hline
\multicolumn{10}{|c|}{$q=5$} \\
\hline
$n/d$ & 1 & 2 & 3 & 4 & 5 & 6 & 7 & 8 & 9 \\
\hline
5 & 25 & 25-30 & 25-30 & 19-25$^{*}$ & & & & & \\
\hline
6 & 25$^a$ & 25-51 & 25-51 & 19-25 & 15-25$^{*}$ & & & & \\
\hline
7 & 29-34 & 25-66 & 25-81 & 19-57$^{t6}$ & 25-34 & 12-15$^{*}$ & & & \\
\hline
8 & 33-40 & 29-75 & 25-88 & 19-65$^{t6}$ & 22-65$^{t6}$ & 26-40 & 10$^{*}$ & & \\
\hline
9 & 37-43 & 33-83 & 25-130 & 21-73$^{t6}$ & 22-73$^{t6}$ & 26-65 & 25-43 & 8-10$^{*}$ & \\
\hline
10 & 41-45 & 37-114 & 25-177 & 21-81$^{t6}$ & 22-81$^{t6}$ & 26-81$^{t6}$ & 25-49 & 25-45 & 7$^{*}$ \\
\hline
\end{tabular}
}
\end{center}

\medskip

{\bf Acknowledgements.} The first author was partially supported
by the National Scientific Program "Information and Communication
Technologies for a Single Digital Market in Science, Education and
Security (ICTinSES)", financed by the Bulgarian Ministry of
Education and Science. He is also with Technical Faculty, South-Western University, Blagoevgrad, Bulgaria.
The second author was suported by a
Bulgarian NSF contract DN2/02-2016. The research of the third and
forth authors was carried out at the IITP RAS at the expense
of the Russian Fundamental Research Foundation (project No.
19-01-00364). We thank Grigory Kabatiansky for useful discussion concerning the codes under consideration.

\begin{center}
REFERENCES
\end{center}

1.~ {\em  Boyvalenkov P., Delchev K., Zinoviev D. V.,
Zinoviev V. A.,} Codes with two distances: $d$ and $d+1$//
Proceedings of the 16th International Workshop on Algebraic and
Combinatorial Coding Theory, Svetlogorsk (Kaliningrad region.
Russia). 2018.

2.~ {\em   Landjev I.,  Rousseva A., Storme L.,}
On linear codes of almost constant weight and the related
arcs// manuscript, 2019.

3.~ {\em  Bassalygo L. A.,} New upper bounds for error-correcting codes//
Problems of Information Transmission. 1965. V. 1, $N^{\circ}$ 1. P. 41 - 44.

4.~ {\em  Bassalygo L. A., Zinoviev V. A.,} A note on balanced incomplete
block-designs, near-resolvable block-designs, and $q$-ary optimal
constant-weight codes// Problems of Information Transmission.
2017. V. 53. $N^{\circ}$ 1. P. 51-54.

5.~ {\em  Bassalygo L. A., Zinoviev V. A., Lebedev V. S.,}
On $m$-nearly resolvable BIB designs and $q$-ary constant weight codes//
Problems of Information Transmission. 2018. V. 54. $N^{\circ}$ 3. P. 54-61.

6.~{\em Semakov N. V., Zinoviev V. A., Zaitsev V. G.,} Class of maximal
equidistant codes// Problems of Information Transmission. 1969.
V. 5. $N^{\circ}$ 2. P. 84-87.

7.~ {\em  Beth T., Jungnickel D., Lenz H.,} Design Theory.
Cambridge University Press, London, 1986.

8.~ {\em Brouwer A. E.,} Tables of bounds for $q$-ary codes//
 http:www.win.tue.nl/\~{}aeb/

9.~ {\em Conway J. H., Sloane N. J. A.,} Sphere Packings, Lattices
and Groups. Springer -- Verlag, New York, 1988.

10.~ {\em Bogdanova G., Todorov T., Zinoviev V. A.,}  On construction of
$q$-ary equidistant codes// Problems of Information Transmission.
2007. V. 43. $N^{\circ}$ 4. P. 13-36.

11.~ {\em Larman D. G., Rogers C. A., Seidel J. J.,} On two
distance sets in Euclidean space// Bull. London Math. Soc. 1977. V. 9.
P. 261-267.

12.~ {\em Nozaki H.} A generalization of Larman-Rogers-Seidel's
theorem// Discrete Math. 2011. V. 311 (1011). P. 792-799.

\end{document}